\documentclass[a4paper,fleqn,usenatbib,useAMS]{mnras}

\usepackage{graphicx}	
\usepackage{amsmath}	
\usepackage{amssymb}	
\usepackage{multicol}        
\usepackage{bm}		
\usepackage{pdflscape}	

\usepackage{subfig}

\usepackage{tabulary}
\usepackage{booktabs,longtable,tabularx}
\newcolumntype{L}{>{\raggedright\arraybackslash}X}
\usepackage{ltablex}
\usepackage{caption}
\usepackage{xcolor}

\usepackage{selinput}
\SelectInputMappings{%
  aacute={á},
  ntilde={ñ},
  Euro={€}
}
\usepackage{mathtools}
\DeclarePairedDelimiter\abs{\lvert}{\rvert}%
\makeatletter
\let\oldabs\abs
\def\abs{\@ifstar{\oldabs}{\oldabs*}}

\usepackage[T1]{fontenc}
\usepackage{ae,aecompl}

\usepackage{newtxtext,newtxmath}

\title[Geometrical compression]{Geometrical compression: a new method to enhance the BOSS galaxy bispectrum monopole constraints}

\author[D. Gualdi et al.]
{\parbox{\textwidth}{Davide Gualdi$^{1}$\thanks{Contact e-mail: \href{davide.gualdi.14@ucl.ac.uk}{davide.gualdi.14@ucl.ac.uk}},
Héctor Gil-Marín$^{2,3}$,
Marc Manera$^{4,5}$,
Benjamin Joachimi$^{1}$,
}
\newauthor{
Ofer Lahav$^{1}$}
\vspace{0.4cm}\
\\
\parbox{\textwidth}
{$^{1}$Department of Physics and Astronomy, University College London, Gower Street, London WC1E 6BT, UK \\
$^{2}$Sorbonne Université, Institut Lagrange de Paris (ILP), 98 bis Boulevard Arago, 75014 Paris, France\\
$^{3}$Laboratoire de Physique Nucléaire et de Hautes Energies, Université Pierre et Marie Curie, Paris, France\\
$^{4}$Institut de Física d’Altes Energies (IFAE), The Barcelona Institute of
Science and Technology, Campus UAB, 08193 Bellaterra (Barcelona) Spain\\
$^{5}$Centre for Mathematical Sciences, DAMTP, Cambridge University, Wilberforce Rd, Cambridge CB3 0WA
}}

\date{Last updated 2018 December 6; in original form 2018 August 5}

\pubyear{2018}

\begin{document}
\label{firstpage}
\pagerange{\pageref{firstpage}--\pageref{lastpage}}
\maketitle

\begin{abstract}
We present a novel method to compress galaxy clustering three-point statistics and apply it to redshift space galaxy bispectrum monopole measurements from BOSS DR12 CMASS data considering a $k$-space range of $0.03-0.12\,h/\mathrm{Mpc}$.
The method consists in binning together bispectra evaluated at sets of wave-numbers forming closed triangles with similar geometrical properties: the area, the cosine of the largest angle and the ratio between the cosines of the remaining two angles.
This enables us to increase the number of bispectrum measurements for example by a factor of $23$ over the standard binning (from 116 to 2734 triangles used), which is otherwise limited by the number of mock catalogues available to estimate the covariance matrix needed to derive parameter constraints. 
The $68\%$ credible intervals for the inferred parameters $\left(b_1,b_2,f,\sigma_8\right)$ are thus reduced by $\left(-39\%,-49\%,-29\%,-22\%\right)$, respectively. 
We find very good agreement with the posteriors recently obtained by alternative maximal compression methods. 
This new method does not require the a-priori computation of the data-vector covariance matrix and has the potential to be directly applicable to other three-point statistics (e.g. glaxy clustering, weak gravitational lensing, 21 cm emission line) measured from future surveys such as DESI, Euclid, PFS and SKA.
\end{abstract}





\section{INTRODUCTION}
Three-point (3pt) statistics will be indispensable to fully exploit the large data-sets from current and forthcoming cosmological surveys.
Their most recent applications to galaxy clustering data sets have been on BOSS for both the bispectrum \citep{2017MNRAS.465.1757G} while \citep{2017MNRAS.468.1070S} used the the 3pt correlation function. \citet{2017MNRAS.469.1738S} also measured baryonic acoustic oscillations (BAO) using the 3pt correlation function and \citet{2018MNRAS.478.4500P} detected them using the bispectrum. 
For the 21cm emission line, 3pt statistics have been investigated by \citet{2018arXiv180202578H}.

Weak-lensing 3pt statistics in Fourier and real space have also been explored (\citealp{2004MNRAS.348..897T,2009A&A...508.1193J,2013MNRAS.429..344K,2013arXiv1306.4684K,2005A&A...431....9S}), including early applications to data \citep{2005A&A...442...69K,2014MNRAS.441.2725F}.

As shown also recently by \citet{2018arXiv180707076Y}, considering the bispectrum together with the power spectrum significantly improves the constraints on cosmological parameters, even if using only the bispectrum monopole severely limits these improvements. In order to include higher multipoles, from the data analysis side, data-vector compression becomes essential.

In previous work we introduced two compression methods for the redshift space galaxy bispectrum in \citet{2018MNRAS.tmp..252G}, Paper I hereafter, and tested them on bispectrum monopole measurements from BOSS DR12 data in \citet{Gualdi2018:1806.02853v1}, Paper II hereafter. Both methods are variations of the method presented in \citet{2000MNRAS.317..965H} and named MOPED, which achieves maximal compression of the original data-vector by extending to the multiple parameters case the Karhunen-Loève algorithm first introduced in \citet{1997ApJ...480...22T}. The two techniques require an approximate analytic expression for the data-vector covariance matrix. These methods compress the original data-vector to a new one with dimension corresponding to the number of model parameters constrained, hence the name "maximal" compression. Since in Paper II we have shown that the two methods produce very similar results, for the sake of clarity in this work we compare the new method to just one of them. In particular we use the method consisting in running a Markov Chain Monte Carlo sampling (MCMC) on the compressed data-vector, labelling these results "maximal compression".

Here we present a new compression method which consists in averaging bispectra triangle configurations of wave-numbers that have similar geometrical properties. In order to derive parameter constraints, we again use MCMC sampling on the compressed data-vector. We label the method geometrical compression (MC-GC).

In particular we define new bins in terms of the triangle configurations area and functions of the internal angles. The area parametrises the physical scales information encoded in the two power spectra products present in the bispectrum analytic expression. At the same time the angles are the variables on which it depends the value of the second order perturbation theory kernel. This can be seen in Figure 2 of \citep{2017MNRAS.465.1757G} where the oscillating pattern in the bispectrum repeats itself because the angles are the same, even if the sizes of k1,k2,k3 increase.
Therefore using these parameters to compress the bispectrum proves to be much more optimal than simply using larger bins defined in terms of the triangle configurations sides.

In Sec. \ref{sec:data_mocks_analysis} the data set and the galaxy mocks used together with the settings of our analysis are described. In Sec. \ref{sec:datav_cov} we present the analytical model used for the considered data-vector.
Sec. \ref{sec:new_tr_par} introduces the transformation through which we compress the original data-vector. Sec. \ref{sec:bins_choice} describes how to optimally choose the number of bins for the new parameters characterising the compressed data-vector. In Sec. \ref{sec:gc_mc_kl_pc} we compare the MC-GC results with the ones from standard MCMC and one of the two alternative maximal compression techniques, described in Paper II.
In Sec. \ref{sec:conclusions} we conclude and discuss potential future extensions.

\section{DATA, MOCKS AND ANALYSIS}
\label{sec:data_mocks_analysis}
The power spectrum monopole, quadrupole and bispectrum monopole have been measured from the DR12 CMASS sample ($0.43\leq z \leq 0.70$) of the Baryon Oscillation Spectroscopic Survey (BOSS, \citealp{2013AJ....145...10D}) which is part of the Sloan Digital Sky Survey III \citep{2011AJ....142...72E}. For more details see \citet{2017MNRAS.465.1757G} and \citet{2017MNRAS.470.2617A}.

The covariance matrix used to estimate the cosmological parameters of interest via standard MCMC on the full data-vector has been numerically estimated using 1400 of the 2048 galaxy catalogues of the MultiDark Patchy BOSS DR12 mocks by \citet{2016MNRAS.456.4156K}.
We only use 700 when the compressed data-vector is used in order to consistently compare the new method here presented with the results obtained in Paper II.
The underlying cosmology used to realise these mocks is: $\Omega_{\Lambda}(z=0) = 0.692885$, $\Omega_{\mathrm{m}}(z=0) = 0.307115$, $\Omega_{\mathrm{b}}(z=0) = 0.048$, $\sigma_8=0.8288$, $n_{\mathrm{s}} = 0.96$, $h_0 = 0.6777$.

\label{sec:analysis_settings}
We fix the bin size for the power spectrum monopole and quadrupole to $\Delta k = 0.01 h/\mathrm{Mpc}$.
We estimated the bispectrum monopole from both data and mocks using different multiples of the fundamental frequency defined as $k_f^3 = \frac{(2\pi)^3}{V_{\mathrm{s}}}$ where $V_{\mathrm{s}}$ is the survey volume. $V_{\mathrm{s}}$ has been set to the mocks case value, which was the one of a cubic box volume $V_{\mathrm{s}} = L_{\mathrm{b}}^3 = (3500\, \mathrm{Mpc}/h)^3$. 

For the bispectrum we considered the bin sizes $\Delta k_{6,5,2} =6,5,2\, \times\, k_f$ respectively, corresponding to 116 and 2734 triangles used between $0.02<k_i\, [h/\mathrm{Mpc}] < 0.12$. The largest bin size $\Delta k_6$ corresponds to the one used in the standard BOSS analysis performed by \citet{2017MNRAS.465.1757G}.

For the same reason, we use the same range of scales:  $k_{\mathrm{min}} = 0.03\, h/\mathrm{Mpc}$ and $k_{\mathrm{max}} = 0.09\, h/\mathrm{Mpc}$ for both power spectrum monopole and quadrupole, $k_{\mathrm{min}} = 0.02\, h/\mathrm{Mpc}$ and $k_{\mathrm{max}} = 0.12\, h/\mathrm{Mpc}$ for the bispectrum monopole. 

The fiducial cosmology chosen for the analysis corresponds to a flat $\Lambda$CDM model similar to the one reported in
\citet{2016A&A...594A..13P} and recently in \citet{2018arXiv180706209P}. In particular we set $\Omega_{\mathrm{m}}(z=0) = 0.31$, $\Omega_{\mathrm{b}}(z=0) = 0.049$, $A_{\mathrm{\mathrm{s}}}=2.21\times 10^{-9}$, $n_{\mathrm{s}} = 0.9624$, $h_0 = 0.6711$.
As in Paper II, in order to compute the numerical derivatives of the data-vector with respect to the model parameters, we fixed the fiducial value of the bias model parameters, the growth rate and the amplitude of dark matter oscillations to the ones obtained by running a preliminary low resolution MCMC ($b_1 =2.5478$, $b_2=1.2127$, $f =0.7202$, $\sigma_8 = 0.4722$).

Since for the range of scales considered (quasi-linear regime), the Fingers-of-God parameters for both power spectrum and bispectrum were compatible with zero, $\sigma^{\mathrm{FoG}}_{\mathrm{B}_k} $ and $\sigma^{\mathrm{FoG}}_{\mathrm{P}_k}$ have been set to zero.
In Paper II we tested that the choice of fiducial parameters used to compute the analytical covariance matrix and the derivatives of the mean of the data-vector does not significantly influence the results of the compression.
\section{DATA-VECTOR}
\label{sec:datav_cov}
 We use the estimators described in \cite{2015MNRAS.451..539G} and \citet{2017MNRAS.465.1757G} to measure the power spectrum monopole and quadrupole together with the bispectrum monopole from the data and the galaxy catalogues. 
 In this work we constrain the model parameters using the joint data-vector obtained by combining the power spectrum monopole and quadrupole with the bispectrum monopole.


Almost all the 2pt statistics signal is contained in the first two multipoles of the redshift space galaxy power spectrum, the monopole and the quadrupole ($\ell=0,2$). These can be found by integrating the galaxy power spectrum:

\begin{eqnarray}
\label{pk_02_esp}
 \mathrm{P}_{\mathrm{g}}^{(\ell)}\left(k\right) = \dfrac{2\ell+ 1}{2}\int^{+1}_{-1}d\mu\,\mathrm{P}_{\mathrm{g}}^{(s)}\left(k, \mu\right)L_{\ell}\left(\mu\right)\,,
\end{eqnarray}

\noindent where $L_{\ell}\left(\mu\right)$ is the $\ell$-order Legendre polynomial and $\mathrm{P}_{\mathrm{g}}^{(s)}\left(k, \mu\right)$ is the redshift space galaxy power spectrum defined in Paper II and originally in the appendix of \citet{2014JCAP...12..029G}.

We adopt the effective model presented in \citet{2014JCAP...12..029G} for the redshift space galaxy bispectrum. This consists in the modification of the redshift space distortions kernels derived from perturbations theory (see the appendix of the paper above for the full expressions). 

The monopole of the bispectrum is obtained by averaging all the possible orientations of a triangle configuration with respect to the line of sight. It can therefore be computed through the integration of the two angular coordinates:

\begin{eqnarray}
\label{bk_0_esp}
\mathrm{B}^{(0)}_{\mathrm{g}}\left(k_1,k_2,k_3\right)
&= \dfrac{1}{4}\int^{1}_{-1} d\mu_1\int^{1}_{-1}d\mu_2 \,\mathrm{B}^{(s)}_{\mathrm{g}}\left(\bm{k}_1,\bm{k}_2,\bm{k}_3\right)
\notag \\
&=\dfrac{1}{4\pi}\int^{1}_{-1} d\mu_1\int^{2\pi}_{0}d\phi \, \mathrm{B}^{(s)}_{\mathrm{g}}\left(\bm{k}_1,\bm{k}_2,\bm{k}_3\right)
\,,
\end{eqnarray}

\noindent where $\mu_i$ is the angle between the $\bm{k}_i$ vector and the line of sight. The angle $\phi$ is defined as $\mu_2\equiv\mu_1x_{12} - \sqrt{1-\mu_1^2}\sqrt{1-x_{12}^2}\cos{\phi}$, where $x_{12}$ is the cosine of the angle between $\bm{k}_1$ and $\bm{k}_2$. More details are given in the appendix of Paper II.

\section{NEW TRIANGLE GEOMETRICAL PARAMETRISATION}
\label{sec:new_tr_par}

We want to regroup the bispectrum data-vector elements in bins defined by different parameters describing the triangle configurations. The idea underlying this procedure is that similar triangular shapes will result in similar sensitivity to the cosmological parameters. This is because the perturbation kernels depend in particular on the cosine of the angles between the sides of the triangle.

Given the three triangle sides $\left(k_1,k_2,k_3\right)$ normally characterising an element of the redshift space galaxy bispectrum monopole data-vector, we define three new variables. The first is the square root of the area of the triangle, which we label $\aleph$ ("aleph"). It can be computed using Heron's formula:

\begin{eqnarray}
A = \sqrt{s(s-k_1)(s-k_2)(s-k_3)} \quad \Longrightarrow \quad \aleph \equiv \sqrt{A},
\end{eqnarray}

\noindent where $s = \frac{1}{2} (k_1+k_2+k_3)$ is the semi-perimeter of the triangle. The $\aleph$ parameter keeps track of the physical scales probed by the triangle configuration. Therefore $\aleph$ is a variable that takes encodes the information the two linear power spectra present in the bispectrum tree-level expression (see Paper I or II appendixes for the explicit expression).

The second variable which we use to characterise a triangle is the cosine of the largest angle\footnote{In this case we mean the "interior" angle of the triangles, which differs from the angles between $k$-vectors used in the perturbation theory kernels by a factor of $\pi$, since the sum of $k$-vectors must be equal to zero.}, $\daleth = \cos\psi_{\mathrm{max}}$ (pronounced "daleth"). This choice allows one to describe whether the triangle is acute or obtuse. If $  \cos(\pi/3) = 1/2  > \daleth > 0 $ the triangle is acute. In this case either the three sides are all approximately the same or two of them are larger than a third one. If $ -1 < \daleth < 0 $ the triangle is obtuse. The triangle could then have either a side much larger than the other two (the one opposite to $\psi_{\mathrm{max}}$) or two sides of similar length with a third smaller one.
In order to distinguish between the pair of possibilities above described, as a third variable we consider the ratio between the cosines of the intermediate and smallest angles, $\gimel = \cos\psi_{\mathrm{int}}/\cos\psi_{\mathrm{min}}$ (pronounced "gimel"). All the cosines can be computed using the cosine rule for a triangle

\begin{eqnarray}
k_l^2 = k_m^2 + k_n^2 - 2k_m k_n\cos\psi_{mn},
\end{eqnarray}

\noindent where $\cos\psi_{mn}$ is the angle between the triangle sides $k_m$ and $k_n$. 
The variables $\daleth,\gimel$ encode the geometrical information strongly affecting the value of the second order perturbation theory kernel present in the bispectrum expression. These variables allow to regroup together triangle configurations returning similar kernel values because of the similar geometrical properties.
Therefore each triangle configuration can be described as a function of the three variables $(\aleph,\daleth,\gimel)$ and the same is true for each bispectrum monopole data-vector element

\begin{eqnarray}
\mathrm{B}^{(0)}_{\mathrm{g}}\left(k_1,k_2,k_3\right) \quad \Longrightarrow \quad \mathrm{B}^{(0)}_{\mathrm{g}}\left(\aleph,\daleth,\gimel\right) . 
\end{eqnarray}

\noindent The vice-versa relation is also valid, each set $\left(\aleph,\daleth,\gimel\right)$ corresponds to a triangle configuration described by a choice of the three sides $\left(k_1,k_2,k_3\right)$. The compression consists in using large enough bins for the new variables $\left(\aleph,\daleth,\gimel\right)$ so that the bispectra of triangles with similar geometrical properties contribute to the same new data-vector element. Once the coordinate conversion has been done for all the triangle configurations, the binning for the new coordinates can be defined by finding the minimum and maximum values for the new parameters $(\aleph,\daleth,\gimel)$. Given a choice for the number of bins for each new coordinate $(n_{\aleph}, n_{\daleth}, n_{\gimel})$, the potential dimension of the new data-vector is $n_{\aleph} \times n_{\daleth} \times n_{\gimel}$. However, as is the case when using the three sides $(k_1, k_2,k_3)$ to describe the triangle, several combinations of $(\aleph,\daleth,\gimel)$ actually do not satisfy the triangle inequalities, and therefore no triplet $(k_1, k_2,k_3)$ will contribute to that particular bin. Moreover even if a particular combination of  $(\aleph,\daleth,\gimel)_k$ does represent a triangle configuration, it is not certain that the triangle bin defined by $(\aleph,\daleth,\gimel)_k$ will contain modes since the original number of triangles in $(k_1, k_2,k_3)$ coordinates was finite. The new data-vector $\bm{g}$ is obtained by averaging over all the bispectra in the non-empty triangle sets defined by different combinations of the coordinates $(\aleph,\daleth,\gimel)$:

\begin{eqnarray}
\label{eq:algorithm}
g_k(\aleph, \daleth, \gimel)_k = \dfrac{1}{N^{\mathrm{tr.}}_i}
\sum^{N^{\mathrm{tr.}}_i}_{j\,:\,(k_1,k_2,k_3)_j\in(\aleph, \daleth, \gimel)_k}
\mathrm{B}^{(0)}_{\mathrm{g}}(k_1,k_2,k_3)_j\,,
\end{eqnarray}

\noindent where each new data-vector element has been normalised by dividing by the number of triangles belonging to the same set defined by a particular combination of $(\aleph,\daleth,\gimel)_k$, $N^{\mathrm{tr.}}_k$. 

\begin{table}
	\centering
	\caption[Best-fit parameters]{
	Best-fit parameters. 
    Mean values of the posterior distributions and $68\%$ credible intervals for the MCMC sampling on the full data vector, the "maximal" and the MC-GC compression methods. 
	The largest $k$-binning $\Delta k_6$,
	the size used in the BOSS analysis, corresponds to the lowest number of triangles (116). For it we show the best-fit parameters obtained via MCMC sampling using the full data-vector. For the thinnest binning $\Delta k_2$, corresponding to the highest number of triangles (2734), we compare the three compression methods. The results shown for the MC-GC method are relative to the cases with $N_{\bm{g}}^{\mathrm{max.}}=196$ (orange) and $N_{\bm{g}}^{\mathrm{max.}}=117$ (yellow). The observed shift in the mean values as a function of the number of considered triangles is due to the strong degeneracy present between the model parameters which gets partially lifted when, thanks to the compression, more triangle configurations are considered.}
	\label{tab:best_fit_par_ch5}
  \begin{tabular}{ *{5}{c} }
    \toprule
            & \multicolumn{1}{c}{$\Delta k_6$} & \multicolumn{3}{c}{$\Delta k_2$} \\
    \cmidrule(lr){2-2}\cmidrule(lr){3-5}
            &  MCMC  & Max. Comp.  & \multicolumn{2}{c}{MC-GC}\\
    \cmidrule(lr){2-2}\cmidrule(lr){3-5}
    $b_1$        &2.41$\,\pm\,$0.22 & 2.33$\,\pm\,$\colorbox{-red!75!green}{0.14} &  2.25$\,\pm\,$\colorbox{yellow!70}{0.15}     &
    2.22$\,\pm\,$\colorbox{yellow!65!red!70!}{0.14}\\
    $b_2$        &1.00$\,\pm\,$0.40 & 0.72$\,\pm\,$\colorbox{-red!75!green}{0.22} &  0.64$\,\pm\,$\colorbox{yellow!70}{0.25}     &
    0.68$\,\pm\,$\colorbox{yellow!65!red!70!}{0.21}\\
    $f  $        &0.69$\,\pm\,$0.08 & 0.63$\,\pm\,$\colorbox{-red!75!green}{0.06} &  0.64$\,\pm\,$\colorbox{yellow!70}{0.06}     &
    0.65$\,\pm\,$\colorbox{yellow!65!red!70!}{0.06}\\
    $\sigma_8$   &0.50$\,\pm\,$0.04 & 0.53$\,\pm\,$\colorbox{-red!75!green}{0.03} &      0.52$\,\pm\,$\colorbox{yellow!70}{0.04}     &
    0.53$\,\pm\,$\colorbox{yellow!65!red!70!}{0.03}\\
    \bottomrule
    \end{tabular}
\end{table}


\setlength{\tabcolsep}{4pt}
\begin{table}
	\centering
	  \caption[Improvement in parameter constraints]{ Improvement in parameter constraints shown are the relative change of the $68\%$ credible intervals for the $\Delta k_2$  $k$-binning case with respect to the $\Delta k_6$ and $\Delta k_5$ (green) cases. For the MC-GC method we show two cases, for $N_{\bm{g}}^{\mathrm{max.}}=196$ (orange) and $N_{\bm{g}}^{\mathrm{max.}}=117$ (yellow). MC-GC obtains very similar improvements, in terms of tighter parameter constraints, to the ones obtained via maximal compression. Notice the substantial difference in parameter constraints improvements between the standard MCMC case using 195 triangles and the MC-GC case recombining 2734 triangles into 194 new data-vector elements.}
	\label{tab:improvements_ch5}
  \begin{tabular}{ *{6 }{c} }
\toprule
& $\Delta\theta^{\mathrm{mc}}_{\Delta k_6}$ 
& \multicolumn{4}{c}{$\dfrac{\Delta\theta^{\mathrm{comp.}} - \Delta\theta^{\mathrm{mc}}_{\Delta k_6}}{\Delta\theta^{\mathrm{mc}}_{\Delta k_6}}\;\left[\%\right]$}  \\
\cmidrule(lr){2-2}\cmidrule(lr){3-6}
        &  MCMC  & MCMC & Max. Comp.  & \multicolumn{2}{c}{MC-GC} \\
\cmidrule(lr){2-2}\cmidrule(lr){3-6}
        &  $N_{\mathrm{tr}}=116$  & $N_{\mathrm{tr}}=195$ & $N_{\mathrm{el.}}=4$ &$N_{\bm{g}}=115$ &$N_{\bm{g}}=194$ \\
\cmidrule(lr){2-2}\cmidrule(lr){3-6}
$\Delta b_1 $     & 0.22 & \colorbox{red!20!blue!20!green!70!}{-23.9} & \colorbox{-red!75!green}{-37.1} &  \colorbox{yellow!70}{-30.5} & \colorbox{yellow!65!red!70!}{-38.8}\\
$\Delta b_2 $     & 0.40 & \colorbox{red!20!blue!20!green!70!}{-34.9} & \colorbox{-red!75!green}{-46.1} &  \colorbox{yellow!70}{-39.3} & \colorbox{yellow!65!red!70!}{-48.6}\\
$\Delta f   $     & 0.08 & \colorbox{red!20!blue!20!green!70!}{-18.5} & \colorbox{-red!75!green}{-27.8} &  \colorbox{yellow!70}{-24.4} & \colorbox{yellow!65!red!70!}{-29.1}\\
$\Delta \sigma_8$ & 0.04 & \colorbox{red!20!blue!20!green!70!}{-12.9} & \colorbox{-red!75!green}{-22.8} &  \colorbox{yellow!70}{-16.6} & \colorbox{yellow!65!red!70!}{-22.0}\\
\cmidrule(lr){1-6}
\multicolumn{2}{c}{$\Big\langle\dfrac{\Delta\theta
- \Delta\theta^{\mathrm{mc}}_{\Delta k_6}}{\Delta\theta^{\mathrm{mc}}_{\Delta k_6}}\;\left[\%\right]\Big\rangle$}  &-22.5 &-33.5 & -27.7 & -34.1\\

\bottomrule
    \end{tabular}
\end{table}

\begin{figure}
\centering
\includegraphics[width=1.025\linewidth]
{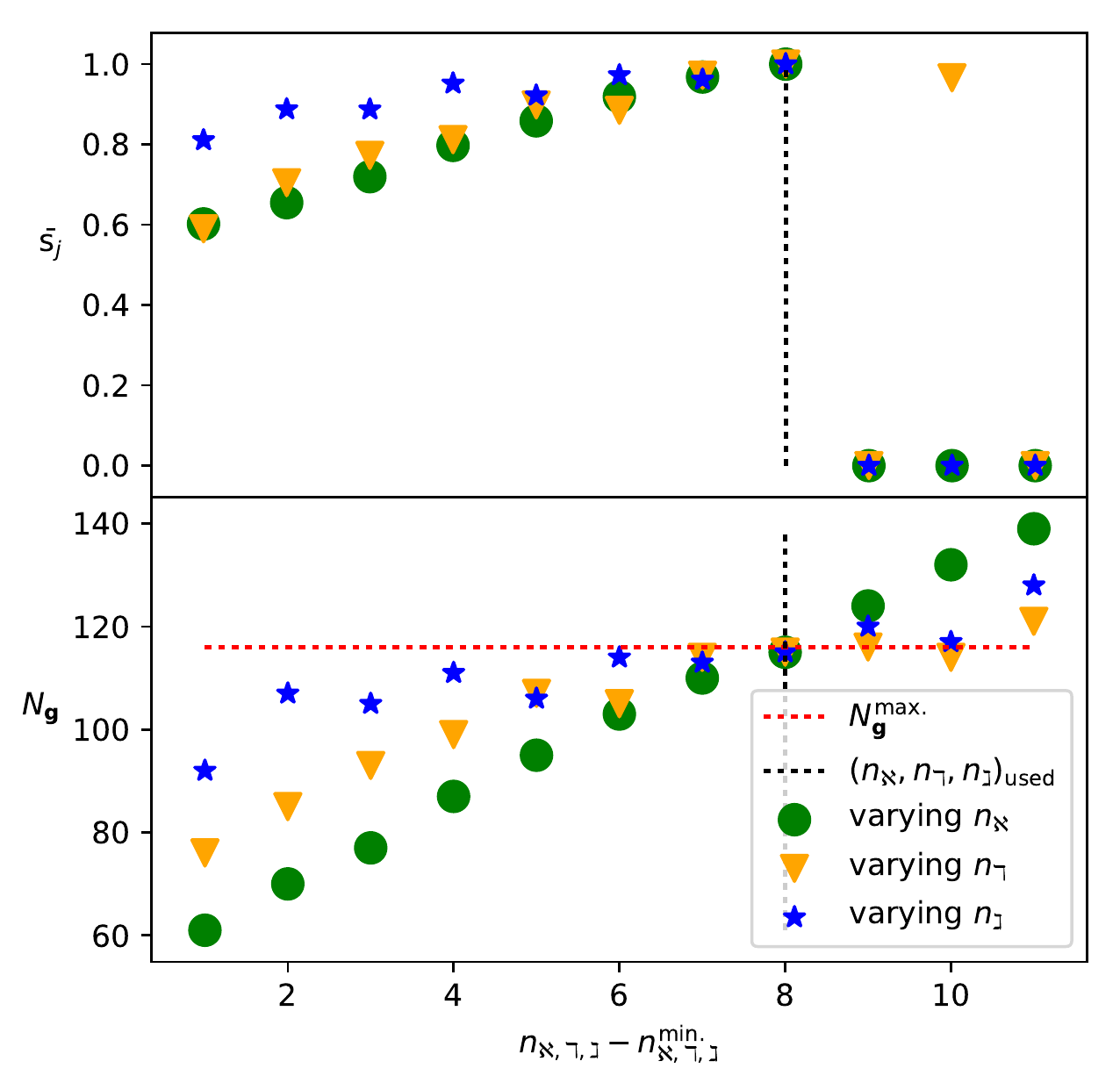}
\caption{Variation of the parameter $\bar{\mathrm{s}}_{j}$ in the $N_{\bm{g}}^{\mathrm{max}} = 117$  case, used to choose the number of bins for the new parameters, and of the number of element of the new data-vector $N_{\bm{g}}$ as a function of $(n_{\aleph}, n_{\daleth}, n_{\gimel})$. $n_{\aleph,\daleth,\gimel}^{\mathrm{min.}}$ is a normalisation on the $x$-axis used to show the same number of different configurations, obtained by varying one of the bins numbers $(n_{\aleph}, n_{\daleth}, n_{\gimel})$ while keeping the other fixed to the optimal value, on both left and right sides of the optimal set $(n_{\aleph}, n_{\daleth}, n_{\gimel})_{\mathrm{opt.}}$. In particular for the case shown we used $(n_{\aleph}^{\mathrm{min.}}=3, n_{\daleth}^{\mathrm{min.}}=2, n_{\gimel}^{\mathrm{min.}}=12)$.  The horizontal red line shows the imposed upper limit to the number of new data-vector elements, $N_{\bm{g}}^{\mathrm{max.}}$.The vertical black line indicates the chosen set of $(n_{\aleph}, n_{\daleth}, n_{\gimel})$ for which $\mathrm{s}_{j}$ was the highest for $N_{\bm{g}}<N_{\bm{g}}^{\mathrm{max.}}$.}
\label{fig:varying_sj}
\end{figure}

\section{NUMBER OF BINS: OPTIMAL CHOICE}
\label{sec:bins_choice}
For the construction of the new data-vector it is necessary to define how many bins will be used to divide the range of each parameter. In order to optimise the choice of these three numbers, $(n_{\aleph}, n_{\daleth}, n_{\gimel})$, we suggest the following procedure. The idea is to "sample" the sensitivity of the new data-vector to the considered model parameters for the different choices of $(n_{\aleph}, n_{\daleth}, n_{\gimel})$. The most straightforward way to do so is to consider the derivatives of the data-vector model with respect to the parameters. These can be computed assuming a fiducial cosmology which in our case was described in Sec \ref{sec:analysis_settings}. 

In order to transform the derivatives of the standard bispectrum monopole data-vector into the derivatives of the new one, it is sufficient to apply the same algorithm used to convert the bispectrum into $\bm{g}$ given in Eq. \ref{eq:algorithm}, because the transformation is linear. At this point we have a list of $\bm{g}_{,i} = \partial\bm{g}/\partial\theta_i$ for all the elements of the model parameter vector $\bm{\theta}$. The target is to combine these vectors into a unique number expressing the sensitivity of the new data-vector $\bm{g}$ for a certain choice of $(n_{\aleph}, n_{\daleth}, n_{\gimel})$. We call $N_{\bm{g}}$ the dimension of the new data-vector $\bm{g}$ and $N_{\bm{\theta}}$ the number of model parameters. $N_{\bm{g}}$ is of course a function of the number of bins of the new coordinates, $N_{\bm{g}}(n_{\aleph}, n_{\daleth}, n_{\gimel})$. For each of the model parameters $\theta_i$ and for a particular choice of the number of bins $(n_{\aleph}, n_{\daleth}, n_{\gimel})_j$ we derive a single number defined as

\begin{eqnarray}
\mathrm{S}_{ij} \equiv \sum_{k=1}^{N_{\bm{g}}(n_{\aleph}, n_{\daleth}, n_{\gimel})_j} \,\dfrac{1}{N_k^\mathrm{tr.}}\, \abs{\dfrac{\partial g_k}{\partial\theta_i}}.
\end{eqnarray}

\noindent $\mathrm{S}_{ij}$ is a proxy for the sensitivity of the new data-vector $\bm{g}$ defined for a particular choice of number of bins $(n_{\aleph}, n_{\daleth}, n_{\gimel})_j$ with respect to variations of the model parameter $\theta_i$. Notice that each term of the sum, before being added, is normalised by the number of triangles regrouped in the new bin defined by a set of coordinates $(\aleph,\daleth,\gimel)_k$.

The next step consists of combining these proxies for all the model parameters. This in order to obtain a single number describing the overall sensitivity of $\bm{g}$ for a determinate choice of $(n_{\aleph}, n_{\daleth}, n_{\gimel})_j$. We then normalise each $i$-th $\mathrm{S}_{ij}$ dividing by the maximum value of $\mathrm{S}_{ij}$ for all the possible $(n_{\aleph}, n_{\daleth}, n_{\gimel})_j$ combinations

\begin{eqnarray}
\mathrm{s}_{ij} \equiv \dfrac{\mathrm{S}_{ij}}{\mathrm{max}\left[\mathrm{S}_{ij}\right]_{\forall j}},
\end{eqnarray}

\noindent so that for all $\theta_i$ then $0< \mathrm{s}_{ij} \leq 1$. Finally, all the $N_{\bm{\theta}}$ $\mathrm{s}_{ij}$ for each $(n_{\aleph}, n_{\daleth}, n_{\gimel})_j$ combination can be merged into a unique number by doing

\begin{eqnarray}
\bar{\mathrm{s}}_{j} \equiv \sum_{i = 1}^{N_{\bm{\theta}}}\, \mathrm{s}_{ij}.
\end{eqnarray}
\noindent We consider $\bar{\mathrm{s}}_{j}$ as the proxy encoding the overall sensitivity to the model parameters of the new data-vector $\bm{g}$, defined by a particular choice of the triplet $(n_{\aleph}, n_{\daleth}, n_{\gimel})_j$. 
Since we may want to limit the dimension of $\bm{g}$ in the algorithm, we include a condition setting $\bar{\mathrm{s}}_{j} = 0$ when $N_{\bm{g}}(n_{\aleph}, n_{\daleth}, n_{\gimel})_j \geq N_{\bm{g}}^{\mathrm{max}}$.
The standard BOSS analysis bispectrum data-vector, limited to the range of scales that we consider, has 116 triangles ($\Delta k_6$ binning case defined in Section \ref{sec:analysis_settings}). We use the measurements done for the $\Delta k_2$ binning case corresponding to 2734 triangles for the bispectrum monopole.

We consider two cases, $N_{\bm{g}}^{\mathrm{max}} = 117$ and $N_{\bm{g}}^{\mathrm{max}} = 196$, compressing the original bispectrum monopole by a factor of $\sim23$ and $\sim14$, respectively.
The $N_{\bm{g}}^{\mathrm{max}} = 196$ is used to study the difference between MC-GC  and the standard MCMC on the full data-vector given by the $\Delta k_5$ binning of the triangle sides.

For $N_{\bm{g}}^{\mathrm{max}} = 117$, $\bar{\mathrm{s}}_{j}$ has been computed for all the $(n_{\aleph}, n_{\daleth}, n_{\gimel})_j$ combinations with $1 \leq n_{\aleph}, n_{\daleth}, n_{\gimel} \leq 25$. 
With these settings we obtained the highest value for $\bar{\mathrm{s}}_{j}$ in the case of $(n_{\aleph} = 10,\, n_{\daleth} = 9,\, n_{\gimel} =19)$ corresponding to a dimension $N_{\bm{g}}(10, 9, 19) = 115$. 
For $N_{\bm{g}}^{\mathrm{max}} = 196$, $\bar{\mathrm{s}}_{j}$ has been computed for all the $(n_{\aleph}, n_{\daleth}, n_{\gimel})_j$ combinations with $5 \leq n_{\aleph}, n_{\daleth}, n_{\gimel} \leq 30$. 
With these settings we obtained the highest value for $\bar{\mathrm{s}}_{j}$ in the case of $(n_{\aleph} = 22,\, n_{\daleth} = 10,\, n_{\gimel} =16)$ corresponding to a dimension $N_{\bm{g}}(22,10,16) = 194$. 
Figure \ref{fig:varying_sj} shows the variation of $\bar{\mathrm{s}}_{j}$ as function of each number of bins for the $N_{\bm{g}}^{\mathrm{max}} = 117$ case, keeping the others fixed to the optimal value. In the last two columns of Table \ref{tab:best_fit_par_ch5} and Table \ref{tab:improvements_ch5} we show that the difference between the mean of the 1D posterior distributions obtained for the two cases $N_{\bm{g}}^{\mathrm{max}} = 117$ and $N_{\bm{g}}^{\mathrm{max}} = 196$ is small and that improvement on parameter constraints are similar. 


\begin{figure}%
    \centering
    \includegraphics[width=0.5\textwidth]
    {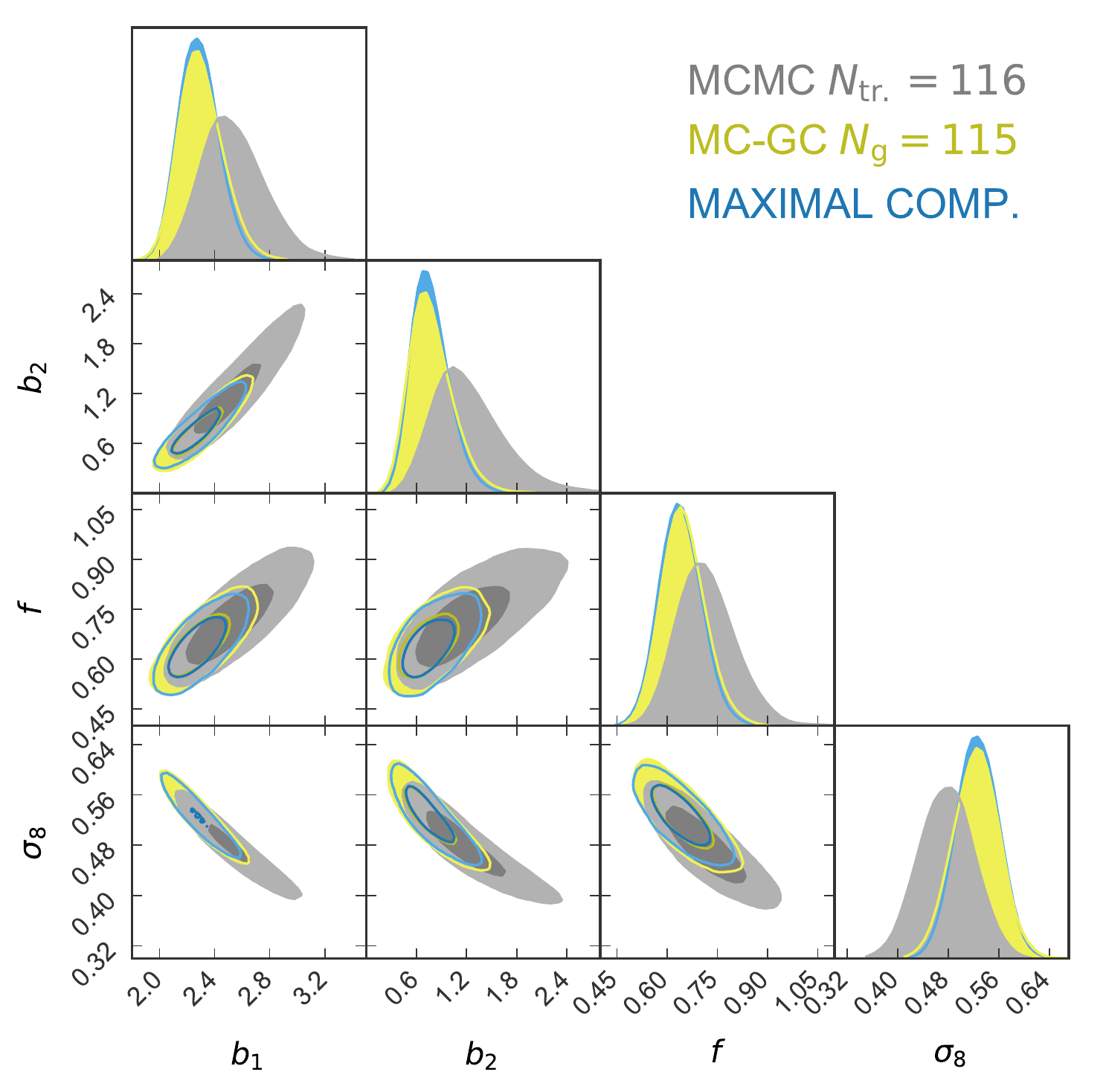}
    \caption{
    Compression performance: 2-D  $68\%$ and $95\%$ credible regions are shown respectively for the $\Delta k_6$ standard MCMC sampling (MCMC in grey, 116 triangles), $\Delta k_2$ maximal compression (MCMC on the compressed data vector in blue, obtained using the maximal compression method presented in \citet{2018MNRAS.tmp..252G} on the original 2734 triangles) and  $\Delta k_2$ geometrical compression (MC-GC in yellow, 2734 triangles) cases. The MC-GC case shown is obtained imposing that the dimension of the compressed data-vector satisfies $N_{\bm{g}}^{\mathrm{max.}}=116$. The agreement between maximal compression and MC-GC posterior distributions is remarkable. Without the need of an analytical modelling of the covariance matrix, MC-GC recovers very close posterior distributions to the ones derived using the maximal compression method.
The observed shift between MCMC results using 116 triangles and MC-GC / maximal compression using 2734 triangles is due to the strong degeneracy between the model parameters which is partially lifted when more triangle configurations are used. In particular the shift happens along the degeneration direction of $b_1$, $b_2$ and $f$ with $\sigma_8$ and as described in \citet{Gualdi2018:1806.02853v1} and it may have a statistical origin.
}
    \label{fig:gc_vs_kl_vs_pc}
\end{figure}

\section{COMPRESSION PERFORMANCE}
\label{sec:gc_mc_kl_pc}

We can compare the results obtained via MC-GC ($\Delta k_2$ case) in terms of 1D and 2D the posterior distributions obtained via the standard MCMC sampling ($\Delta k_6$ and $\Delta k_5$ cases) and maximal compression ($\Delta k_2$ case). The comparison is shown in Figure \ref{fig:gc_vs_kl_vs_pc}. Even if it does not need to analytically model the covariance matrix in order to compress the data-vector, MC-GC produces a posterior distribution very close to the one given by the maximal compression method. The agreement is remarkable, especially considering that these compression methods are fairly independent of each other (they have in common only the use of the data-vector derivatives). The precise values of the 1D $68\%$ confidence intervals and of the means of the distribution are reported in Tables \ref{tab:best_fit_par_ch5} and \ref{tab:improvements_ch5}. 

It is important to notice the difference between the MCMC with 116, 195 triangles and the MC-GC results using 115 and 194 combinations of the original 2734 triangles, respectively. It is clear from both Table \ref{tab:improvements_ch5} and Figure \ref{fig:gc_vs_kl_vs_pc} that when the same number of data-vector elements are considered, MC-GC produces much tighter constraints since it is able to exploit the constraining power of the original 2734 triangles. 

The observed shift between MCMC results using 116 triangles and MC-GC / maximal compression using 2734 triangles is due to the strong degeneracy between the model parameters which is partially lifted when more triangle configurations are used in the data-vector.
In terms of time and computing resources, MC-GC is equivalent to standard MCMC sampling and maximal compression method (details given in Paper II).


\section{CONCLUSIONS}
\label{sec:conclusions}
The new compression method presented in this work consists in binning together bispectra evaluated at sets of wave-numbers forming closed triangles with similar geometrical properties: the area, the cosine of the largest angle and the ratio between the cosines of the remaining two angles.

The advantage of the geometrical compression (MC-GC) technique, with respect to maximal compression methods, introduced in \citet{2018MNRAS.tmp..252G} and applied to BOSS data in \citet{Gualdi2018:1806.02853v1}, is that it does not require an analytical modelling of the covariance matrix. This is due to the fact that MC-GC is based on the similarities between the geometrical properties of different triangle configurations and not on their bispectrum values covariance.
In terms of resources and computing time required, these are approximately the same as for the maximal compression method (see Paper II for details), i.e. the time taken by the geometrical compression step is negligible. 
The MC-GC compression is not "maximal" as the ones presented in \citet{2018MNRAS.tmp..252G}. We compressed using MC-GC the bispectrum of 2734 triangle configurations into data-vectors up to $\sim23$ times shorter. 

By compressing the data-vector using the geometrical compression before running the MCMC sampling, we improved BOSS constraints, reducing the $68\%$ credible intervals for the inferred parameters $\left(b_1,b_2,f,\sigma_8\right)$ by $\left(-39\%,-49\%,-29\%,-22\%\right)$, respectively.

Future work will include the development of extensions of the MC-GC method to higher-order statistics, like the trispectrum and tetraspectrum, always using geometrical properties of the $k$-vectors' configurations.
Moreover we are interested in applying MC-GC to weak-lensing and 21 cm emission line 3pt statistics.
Given its immediate and straightforward applicability, we hope that MC-GC will become a standard procedure for future data-sets to study the bispectra and 3pt functions of the cosmological fields of interest. Another interesting point would be to study whether it is possible to efficiently compress 3pt statistics using different geometrical properties of the triangle configurations than the ones used here.

\section*{ACKNOWLEDGMENTS} 
D.G. is supported by the Perren and the IMPACT studentships. 
HGM is supported by Labex ILP (reference ANR-10-LABX-63) part of the Idex SUPER, and received financial state aid managed by the Agence Nationale de la Recherche, as part of the programme Investissements d’avenir under the reference ANR-11-IDEX-0004-02.e.
M.M. acknowledges funding from STFC Consolidated Grants RG84196 and
RG70655 LEAG/506 and has received funding from the European Union’s Horizon 2020 research and innovation programme under Marie Skłodowska-Curie grant agreement No 6655919.
O.L. acknowledges support from a European Research Council Advanced Grant FP7/291329.
C \citep{Kernighan:1988:CPL:576122} and \textsc{python} 2.7 \citep{Rossum:1995:PRM:869369} have been used together with many packages like I\textsc{python} \citep{Perez:2007:ISI:1251563.1251831}, Numpy \citep{DBLP:journals/corr/abs-1102-1523}, Scipy \citep{jones} and Matplotlib \citep{Hunter:2007:MGE:1251563.1251845}. The corner plots have been realised using \textsc{pygtc} developed by \citet{Bocquet2016} 
 



\bibliographystyle{mnras}

\bibliography{ads_reference.bib}


\bsp	
\label{lastpage}
\end{document}